\documentclass[12pt,preprint]{aastex}
\begin{document}

\title{Nodeless differentially rotational Alfv\'en oscillations of
crustal solid-state plasma in quaking neutron star}

\author{
S.I. Bastrukov\altaffilmark{1,3}, H.-K. Chang\altaffilmark{1,2}, I.V. Molodtsova\altaffilmark{3}, and
J. Takata\altaffilmark{4}}

\altaffiltext{1}{Institute of Astronomy,\\
  National Tsing Hua University, Hsinchu, 30013, Taiwan}

\altaffiltext{2}{Department of Physics,\\
  National Tsing Hua University, Hsinchu, 30013, Taiwan}

\altaffiltext{3}{Laboratory of Informational Technologies,\\
Joint Institute for Nuclear Research, 141980 Dubna, Russia}

\altaffiltext{4}{ASIAA/National Tsing Hua University - TIARA, Hsinchu, Taiwan}

\begin{abstract}
  The two-component, core-crust, model of a neutron star with homogenous internal and dipolar external
  magnetic field is studied responding to quake-induced perturbation by substantially nodeless differentially rotational Alfv\'en oscillations of the perfectly conducting crustal matter about axis of fossil magnetic field frozen in the immobile core. 
  The energy variational method of the magneto-solid-mechanical theory
  of a viscoelastic perfectly conducting medium pervaded by magnetic field is utilized to compute the frequency and
  lifetime of nodeless torsional vibrations of crustal solid-state plasma about the dipole magnetic-moment axis of the star. It is found that obtained two-parametric spectral formula for the frequency of this toroidal Alfv\'en mode provides fairly accurate account of rapid oscillations of the X-ray flux during the flare of SGR 1806-20 and SGR 1900+14, supporting the investigated conjecture that these quasi-periodic oscillations owe its origin to axisymmetric torsional oscillations predominately driven by Lorentz force of magnetic field stresses in the finite-depth crustal region of the above magnetars.
  \end{abstract}

Key words: stars: individual (SGR 1900+14, SGR 1806-20); stars: neutron; stars: oscillations

\section{Introduction}
  The quasi-periodic oscillations (QPOs) recently detected in the X-ray flux of flaring SGR 1806-20 and SGR 1900+14 and interpreted as a manifest of quake-induced torsional vibrations of underlying
  magnetar (Israel et al 2005, Watts \& Strohmayer 2006) are among the topical themes
  in current development of the astroseismology of this subclass of neutron stars (e.g. Glampedakis et al 2007; Samuelssen \& Andersson 2007; Levin 2007, van Hoven \& Levin 2008, Lee 2007, 2008; Sotani et al 2007, 2008; Vavoulidis et al 2008)
  whose bursting and quiescent electromagnetic activity is dominated, as is commonly believed, by frozen-in fossil magnetic field of extraordinary intensity (e.g., Woods \& Thompson   2006, Mereghetti 2008). Motivated by this interest, in (Bastrukov et al 2007, 2008, 2009) several possible
  scenarios of post-quake relaxation of magnetar has been investigated with emphasis on regime of
  nodeless torsional vibrations about the dipole magnetic-moment axis of these seismically active neutron stars.
  In particular, in (Bastrukov et al 2007b, 2008), a case of the elastic-force-driven nodeless shear
  oscillations, both torsional --  $_0t_\ell$ and spheroidal -- $_0s_\ell$, locked in the finite-depth crust $\Delta R$ has been studied in some details. It was found that obtained two-parametric spectral equations
  for the frequency of pure elastic spheroidal and torsional modes as a function of multipole degree $\ell$ of nodeless shear oscillations of the form (Bastrukov et al 2008)
 \begin{eqnarray}
 \label{e1.1}
 && \omega(_0s_\ell)=\omega_s(\ell,\omega_e,\lambda),\quad \omega(_0t_\ell)=\omega_t(\ell,\omega_e,\lambda),\quad\quad \ell\geq 1,\\
  \label{e1.2}
 && \omega_e=\frac{c_t}{R},\quad c_t=\sqrt{\frac{\mu}{\rho}},\quad \lambda=1-\frac{\Delta R}{R},\quad \Delta R=R-R_c
\end{eqnarray}
 provide proper account of the low-frequency QPOs frequencies, supporting $\ell$-pole identification
 proposed in works (Watts \& Strohmayer 2007, Samuelssen \& Andersson 2007) and it was
 shown that the unidentified before lowest overtones of QPOs in light-curve of flaring SGR 1806-20 with frequencies 18 and 26 Hz can be interpreted as manifest of dipole, $\ell=1$, spheroidal and torsion nodeless elastic shear vibrations, exhibiting features generic to Goldstone soft modes.

 On the other hand, one can perhaps cast doubt on such interpretation of QPOs as being produced by seismic vibrations driven by solely elastic restoring force, because it rests on a poorly justifiable
 assumption about dynamically passive role of ultra-strong magnetic field defining only direction of axis about which the stellar matter undergo torsional nodeless oscillations and remaining unaltered in the process of post-quake relaxation of magnetar. Instead, it seems more plausible to expect that explosive liberation of magnetic field energy resulting in the crust fracture by relieved magnetic field stresses -- magnetic breaking associated with
 disruption of the magnetic field lines -- must be followed, after the magnetic field lines reconnection,
 by the Lorentz-force-driven Alfv\'en differentially rotational oscillations of the perfectly conducting star
 matter about axis of frozen-in homogeneous magnetic field, either in the entire volume or in the peripheral finite-depth crust.
 This line of argument has been taken into account in the magneto-solid-mechanical model of global torsional nodeless vibrations about the dipole
 magnetic moment axis of neutron star driven by the joint action of Hooke's force of elastic shear stresses
 and Lorentz force of magnetic field stresses (Bastrukov et al 2009). The computed in this latter work the two-parametric frequency spectrum of the form
 \begin{eqnarray}
 \label{e1.3}
 && \omega(_0t_\ell)=\omega_t(\ell,\omega_e,\beta),\quad\quad \beta=\frac{\omega_A^2}{\omega_e^2}=\frac{B^2}{4\pi\mu},\quad\quad \ell\geq 2,\\
 \label{e1.4}
 && \omega_e^2=\frac{c_t^2}{R^2},\quad c_t^2=\frac{\mu}{\rho},\quad\omega_A^2=\frac{v_A^2}{R^2},\quad v^2_A=\frac{B^2}{4\pi\rho}
 \end{eqnarray}
  was found to be consistent with data on QPOs with frequencies $\nu=\omega/2\pi$
  from the range $30\leq \nu \leq 200$ Hz interpreted as being produced by
  these global seismic vibrations of multipole degree $\ell$ from interval $2\leq \ell\leq 20$,
  but unable to explain the above mentioned lowest overtones in the QPOs spectrum of SGR 1806-20.
  Also, from the viewpoint of both outlined models the nature of restoring force responsible for high-frequency QPOs with $\nu=626$ and  $\nu=1837$ Hz remains uncertain.

  In this paper, continuing investigations reported in (Bastrukov et al 2007, 2008, 2009), we focus
  on different, to the above mentioned scenarios, case of post-quake vibrational relaxation of a neutron star.
  Specifically, we consider torsional nodeless Alfv\'en oscillations of the crustal solid-state plasma,
  that is, seismic vibrations restored by solely Lorentz force of magnetic field stresses
  and excited in the peripheral finite-depth seismogenic layer of neutron star model with homogeneous
  internal and dipolar external magnetic field.
  In section 2, a brief outline is given of equations of magneto-solid-mechanics
  appropriate for the perfectly conducting viscoelastic continuous medium pervaded by magnetic field.
  In section 3, the spectral formulae for the frequency and lifetime of this toroidal Alfv\'en vibration mode
  are obtained followed by discussion of inferences that can be made
  from considered scenario of seismic vibrations regarding electrodynamic properties
  of the neutron star matter with homogeneous configuration of the frozen-in magnetic field.
  The relevance of considered asteroseismic model to the detected QPOs in the X-ray flux
  during the flare of SGR 1806-20 and SGR 1900+14 is assessed in section 4. The newly obtained
  results are briefly summarized in section 5.

\section{Governing equations of magneto-solid-dynamics of viscoelastic perfectly conducting
 stellar matter in the presence of frozen-in magnetic field}

  The equation of magneto-solid-dynamics describing mechanical distortions of the
  perfectly conducting continuous medium of extremely high electric conductivity threaded by magnetic field
  are
  \begin{eqnarray}
  \label{e2.1}
  \rho{\ddot u}_i=\nabla_k\,\tau_{ik}+\nabla_k\,\pi_{ik}
 \end{eqnarray}
 where $u_i$ is the field of material displacements - the basic variable of the Newtonian
 elastic dynamics of solid material continuum,
\begin{eqnarray}
 \label{e2.2}
 \tau_{ik}=\frac{1}{4\pi}[B_i\delta B_k+B_k\delta B_i -B_j\delta B_j \delta_{ik}],\quad\delta
 {\bf B}=\nabla\times [{\bf u}\times {\bf B}]
 \end{eqnarray}
 stands for the tensor magnetic field stresses (Franko et al 2000, Bastrukov et al 2009) and
 \begin{eqnarray}
 \label{e2.3}
 \pi_{ik}=2\eta{\dot u}_{ik},\quad {\dot u}_{ik}=\frac{1}{2}[\nabla_i {\dot u}_k+\nabla_k {\dot u}_i]
 \end{eqnarray}
 is the tensor of viscose stresses $\pi_{ik}$ linearly related to the rate of shear strains ${\dot u}_{ik}$
 with $\eta$ being the shear viscosity of stellar matter.
 This last constitutive equation expresses Newtonian law of viscosity which holds for both liquid and solid aggregate
 states of continuous medium.
 The equation of energy balance in the process of compression-free Alfv\'en vibrations highlighting its shear character reads
 \begin{eqnarray}
  \label{e2.4}
 \frac{\partial }{\partial t} \int \frac{\rho {\dot u}^2}{2}\,
 d{\cal V}=-\int \tau_{ik}\,{\dot u}_{ik}\,d{\cal
 V}- \int \pi_{ik}\,{\dot u}_{ik}\,d{\cal V},\quad {\dot u}_{kk}=\nabla_k {\dot u}_k=0.
 \end{eqnarray}
 This equation is obtained by
 scalar multiplication of equation of magneto-solid-mechanics (\ref{e2.1}) with $u_i$ and integration over the star volume. To compute the frequency and lifetime of Alfv\'en oscillations, we take advantage of Rayleigh's energy
 variational method which assumes the following separable representation of fluctuating field of material displacements
 \begin{eqnarray}
 \label{e2.5}
 && u_i({\bf r},t) =a_i({\bf r}){\alpha}(t)
 \end{eqnarray}
 with $a_i({\bf r})$ being the time-independent field of instantaneous displacement and amplitude ${\alpha}(t)$
 describes temporal evolution of fluctuations. The tensors of fluctuating Maxwell's magnetic stresses and Newtonian viscous stresses are too written in similar separable form
 \begin{eqnarray}
  \label{e2.6}
 && \tau_{ik}({\bf r},t)=[{\tilde \tau}_{ik}({\bf r})-\frac{1}{2}{\tilde \tau}_{jj}({\bf r})\delta_{ik}]\alpha(t),\\
  \label{e2.7}
 && {\tilde \tau}_{ik}({\bf r})=\frac{1}{4\pi}
 [B_i({\bf r})\,b_k({\bf r})+B_k({\bf r}),\,\, b_i({\bf r})],\\
 \label{e2.8}
 && b_i({\bf r})=\nabla_k\,[a_i({\bf r})\,B_k({\bf r})-a_k({\bf r})\,B_i({\bf r})],\\
 \label{e2.9}
 && {\pi}_{ik}({\bf r})=2\eta\,a_{ik}({\bf r})\alpha(t),\quad
 a_{ik}({\bf r})=\frac{1}{2}[\nabla_i a_k({\bf r})+\nabla_k a_i({\bf r})].
 \end{eqnarray}
 The idea of above separable representation of fluctuating variables is that on substituting (\ref{e2.5})-(\ref{e2.9}) in (\ref{e2.4}), this latter equation is reduced to equation for $\alpha(t)$ having the well-familiar form of equation of damped oscillations
\begin{eqnarray}
 \label{e2.10}
 &&{\cal M}{\ddot \alpha}+{\cal D}{\dot
 \alpha}+{\cal K}\alpha=0,
 \quad \alpha(t)=\alpha_0\exp(- t/\tau)\cos(\Omega t),\\
 \label{e2.11}
 && \quad \Omega^2=\omega^2\left[1-(\omega\tau)^{-2}\right],\quad
  \omega^2=\frac{{\cal K}}{{\cal M}},\quad \tau=\frac{2{{\cal M}}}{{\cal D}}.
 \end{eqnarray}
 with the integral parameters of the inertia ${\cal  M}$, viscous friction ${\cal  D}$ and stiffness ${\cal  K}$
 given by
 \begin{eqnarray}
  \label{e2.12}
  && {\cal M}=\int \rho({\bf r}) a_i({\bf r})\,a_i({\bf r})\,d{\cal V},\\
 \label{e2.13}
 && {\cal K}_m=\frac{1}{8\pi} \int [B_i({\bf r})\beta_k({\bf r})+B_k({\bf r})\beta_i({\bf r})]\,[\nabla_i a_k({\bf r})+\nabla_k a_i({\bf r})]\,d{\cal V},\\
 \label{e2.14}
  && {\cal D}= \frac{1}{2}\int \eta(r) [\nabla_i a_k({\bf r})+\nabla_k a_i({\bf r})]\,[\nabla_i a_k({\bf r})+\nabla_k a_i({\bf
 r})] \,d{\cal V}.
 \end{eqnarray}
 The above expounded equations are general in the sense that they can be used for computing
 frequency $\omega$ of both the even-parity poloidal and odd-parity toroidal Alfv\'en vibrational modes (Bastrukov et al 1999) and the time of viscous damping of nodeless oscillations in a neutron star models in which
 the density, the magnetic field and the shear viscosity are considered to be arbitrary functions of position.

 As was noted by McDermott et al (1988), the characteristic feature of solid-mechanical treatment of the neutron star asteroseismology, having many features in common with the one adopted in geoseismology (e.g. Lapwood \& Usami 1981, Lay \& Wallace 1995, Aki \& Richards 2003), is that it deals with quake-induced shear vibrations about
 hydrostatic equilibrium  which is presumed to exist, but with no reference to explicit form of equation
 of gravitationally equilibrium state which is normally considered as the free from anisotropic stresses state.
 The anisotropic stresses of different physical nature (elastic, magnetic or gravitational) defining restoring force of shear vibrations are regarded to come into play only after quake-induced perturbation.
 By thus relinquishing the condition of hydrostatic equilibrium of magnetic solid star,
 we gain enormous freedom in the choice of possible configurations of magnetic field inside and outside the star
 with the only constraint that this equilibrium field must obey the equations of magneto-statics.
 The remainder of this paper, which is sequel to (Bastrukov et al 2007b, 2009), considers the neutron star model with homogeneous inside and dipolar outside magnetic field for which this condition is of course met.

 \section{Toroidal  Altv\'en mode of nodeless differentially rotational oscillations in the peripheral finite-depth
 seismogenic crust of a neutron star}

  The application of the magnetic solid star model under consideration to neutron stars rests on the
  common today belief that the frozen in the stellar material magnetic field is fossil (e.g. Flowers \& Ruderman 1977, Bhattacharya \& van den Heuvel 1991, Chanmugam 1992, Spruit 2008). Contrary to the main sequence (MS) stars whose magnetic fields are generated by permanent flows of fluid in the self-exciting dynamo processes, in the neutron stars there are no nuclear energy sources to support convection, so that the fossil magnetic field
  is considered to be frozen in the super dense solid material continuum which before starquake remains immobile
  being highly compressed by immense pressure of self-gravity which is counterbalanced by the degeneracy pressure of Fermi-gas of relativistic electrons in the star crust and non-relativistic neutrons in the core.
  The magnetic field pressure constructively contributes to the total internal pressure, however  its absolute value is much less than the degeneracy pressure, so that magnetic pressure plays insufficient role in hydrostatic equilibrium of pulsars and magnetars\footnote{At normal nuclear density $\rho=n m_n=2.8\, 10^{14}$ g cm$^{-3}$ with
  $m$ being the mass of neutron, the pressure of degenerate Fermi-gas of non-relativistic neutrons $P_F=K_{NR}\,n^{5/3}\approx 10^{33}$ dyn cm$^{-2}$. The typical for magnetars magnetic fields  $10^{14}< B < 10^{16}$ Gauss,  so that the magnetic field pressure, $P_B=B^2/(8\pi)$, falls in the interval $10^{27} < P_M < 10^{31}$, that is, at least two order of magnitude less than $P_F$. For this reason we
  do not take into account the Chandrasekhar-Fermi effect of magnetic flattening of spherical shape
  of star at its magnetic pole which has first been established for homogeneous magnetic field of dipolar external configuration (Chandrasekhar \& Fermi 1953) and confirmed by Ferraro (1954) for a star model with nonhomogeneous poloidal internal and dipolar external field.}, but may play substantial, if not decisive,
  part in the process of their post-quake relaxation (Bastrukov et al 2009).

  In spherical polar coordinates, the components of internal homogeneous field directed along the polar axis are given by
  \begin{eqnarray}
  \label{e3.1}
   && B_r=B\cos\theta,\quad\quad B_\theta=-B\sin\theta,
 \quad\quad B_\phi=0
 \end{eqnarray}
  and external dipolar configuration is described by the field ${\bf B}=\nabla\times {\bf A}$, where
  ${\bf A}=[0, 0, A_\phi={\rm m}_s/r^2]$ is the vector potential with the standard parametrization of the dipole magnetic moment ${\rm m}_s=(1/2)BR^3$ of the star of radius $R$ and by $B$ is
  understood the magnetic field intensity at the magnetic poles of the star, $B=B_p$, as is illustrated
  in Fig.1.

\begin{figure}[h]
\centering{\includegraphics[width=9.cm]{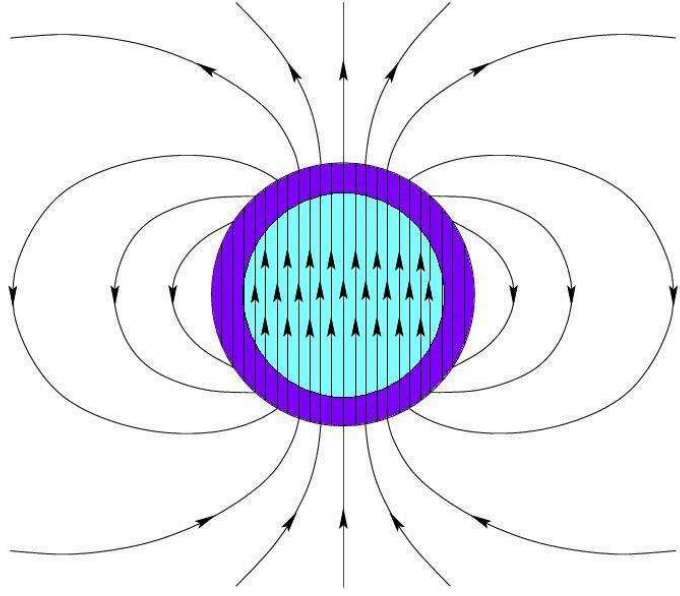}}
\caption{
 The dipolar external and homogenous internal magnetic field of the two-component, core-crust, model of the neutron star under consideration.}
\end{figure}

  The material displacements in the crust undergoing non-radial differentially rotational oscillations
 about magnetic axis of the star are described by the nodeless toroidal field of the form (Bastrukov et al 2007b, 2008b)
 \begin{eqnarray}
 \label{e3.2}
 && {\bf u}({\bf r},t)={\bf a}_t({\bf r})\,\alpha(t),\quad
 {\bf a}_t({\bf r})=[\mbox{\boldmath $\phi$}({\bf r})\times {\bf r}]\\
 \label{e3.3}
 && \mbox{\boldmath $\phi$}({\bf r})=\nabla\chi({\bf r}),\quad \nabla^2\chi({\bf r})=0\\
  \label{e3.4}
 && \chi({\bf r})=[{\cal A}_\ell\,r^\ell+{\cal B}_\ell\,r^{-\ell-1}]\,P_\ell(\zeta),\quad \zeta=\cos\theta
 \end{eqnarray}
 The arbitrary constants ${\cal A}_\ell$ and ${\cal B}_\ell$ are eliminated from the following boundary conditions
 \begin{eqnarray}
   \label{e3.5}
 u_\phi\vert_{r=R_c}=0,\quad
 u_{\phi}\vert_{r=R}=[\mbox{\boldmath $\Phi$}\times {\bf R}]_\phi,\quad
 \mbox{\boldmath $\Phi$}=\nabla_{r=R} P_\ell(\zeta),\quad
  \nabla_{r=R}=\left(0,\frac{\partial }{R\partial
 \theta},\frac{1}{R\sin\theta}\frac{\partial}
 {\partial \phi}\right).
 \end{eqnarray}
 The leftmost equation is the no-slip condition on the core-crust interface, $r=R_c$,
 implying that the amplitude of differentially rotational oscillations
 is gradually decreases as depth of seismogenic layer is increased and turns into zero on the surface of core.
 The condition on the star surface, $r=R$, is dictated by symmetry of the general toroidal solution of the vector Laplace equation as well as the fact that resultant expression of mass parameter of torsional oscillations
 must lead to correct expression for the moment of inertia of a rigidly rotating star (Bastrukov et al 2008). Explicitly, the above boundary conditions are described by coupled algebraic equations
 \begin{eqnarray}
 \label{e3.7}
&&{\cal A}_\ell R_c^{\ell-1}+{\cal B}_\ell R_c^{-\ell-2}=0,\\
\label{e3.8}
&&{\cal A}_\ell R^{\ell}+{\cal B}_\ell R^{-\ell-1}=R
\end{eqnarray}
whose solutions read
\begin{eqnarray}
  \label{e3.9}
 {\cal A}_\ell={\cal N}_\ell,\quad {\cal B}_{\ell}=-{\cal
 N}_\ell\,R_c^{2\ell+1},\quad\quad {\cal
 N}_\ell=\frac{R^{\ell+2}}{R^{2\ell+1}-R_c^{2\ell+1}}.
 \end{eqnarray}
The spherical components of the nodeless toroidal field of instantaneous displacements are given by
\begin{eqnarray}
 \label{e3.10}
 a_{r}=0,\,\, a_{\theta}=0,
 \,\, a_{\phi}=\left[{\cal A}_\ell\,r^\ell+\frac{{\cal B}_\ell}{r^{\ell+1}}\right]
 P^1_\ell(\zeta),\quad P^1_\ell(\zeta)=(1-\zeta^2)^{1/2}\frac{d P_\ell(\zeta)}{d\zeta}.
 \end{eqnarray}
 The snapshot of quadrupole, $\ell=2$, and octupole $\ell=3$, overtones of these axisymmetric differentially rotational vibrations about the dipole magnetic-moment axis of a neutron star is pictured in Fig.2.

\begin{figure}[h]
\centering{\includegraphics[width=9.cm]{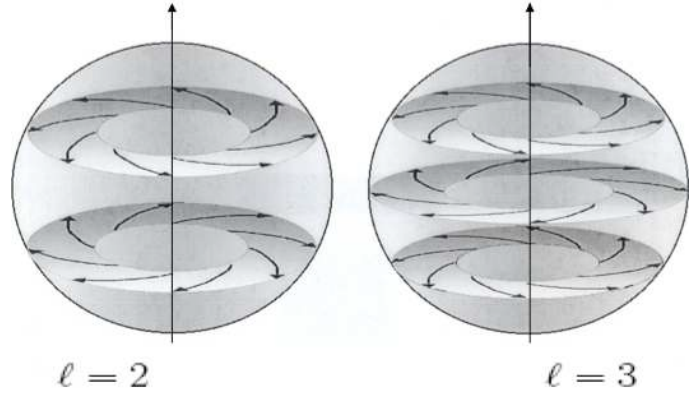}}
 \caption{Quadrupole and octupole overtones of the nodeless differentially rotational, torsional,
 vibrations of crustal matter about the dipole magnetic moment axis of neutron star.}
\end{figure}

 The mass parameter of torsional nodeless vibrations, computed in the approximation of uniform
 density of crustal matter, reads
\begin{eqnarray}
 \label{e3.11}
 {\cal M}&=&4\pi\,\rho R^5\frac{\ell(\ell+1)}{(2\ell+1)(2\ell+3)}\,m_t(\ell,\lambda),\\
 \nonumber
 m_t(\ell,\lambda)&=&(1-\lambda^{2\ell+1})^{-2}\left[1-
 (2\ell+3)\lambda^{2\ell+1}
 + \frac{(2\ell+1)^2}{2\ell-1}\lambda^{2\ell+3}-
 \frac{2\ell+3}{2\ell-1}\lambda^{2(2\ell+1)}\right],\\
 \label{e3.12}
 \lambda&=&\frac{R_c}{R},\qquad 0\leq \lambda <1.
\end{eqnarray}
The integral parameter of viscous friction $D$ is given by
\begin{eqnarray}
 \label{e3.12}
 && {\cal D}(_0t_\ell)=4\pi\eta R^3\frac{\ell(\ell^2-1)}{2\ell+1}
 \,d_t,\quad d_t(\ell,\lambda)={(1-\lambda^{2\ell+1})^{-1}}
 \left[1-\frac{(\ell+2)}{(\ell-1)}\lambda^{2\ell+1}\right].
\end{eqnarray}
For the lifetime of torsional oscillations damped by shear viscosity $\eta$ of crustal material
we obtain
 \begin{eqnarray}
 \label{e3.13}
&& \tau(_0t_\ell)=\frac{2\tau_\nu}{(2\ell+3)(\ell-1)}\,\frac{m_t(\ell,\lambda)}{d_t(\ell,\lambda)}\quad
,\quad \tau_\nu=\frac{R^2}{\nu},\quad \nu=\frac{\eta}{\rho}.
\end{eqnarray}
 In Fig.3, the fractional lifetime
 is plotted as a function of multipole degree $\ell$ of torsional nodeless oscillations
 for indicated values of fractional depths $h$  of crustal seismically active region of the neutron star.

 \begin{figure}[h]
\centering{\includegraphics[width=9.cm]{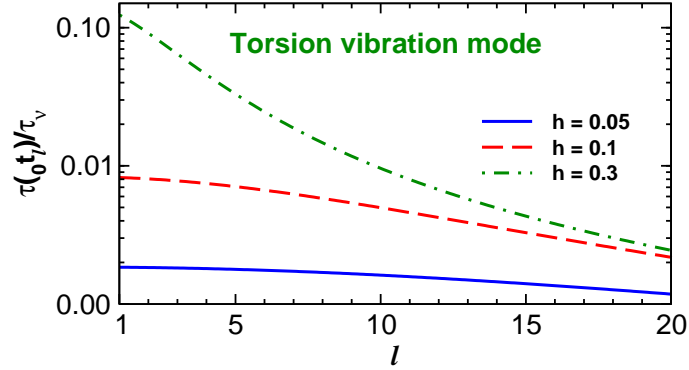}}
 \caption{The fractional lifetime of torsion nodeless oscillations in the neutron star crust as a functions of multipole degree $\ell$ and indicated values of the fractional depth $h$ of peripheral seismogeneic zone.}
\end{figure}

 At this point it may be appropriate to note that replacing, in the last
 equation for ${\cal D}(_0t_\ell)$, the coefficient of shear viscosity $\eta$ by shear
 modulus $\mu$ of crustal matter, the resultant equation becomes identical to the
 expression for stiffness $K_e(_0t_\ell)$ of the elastic-force-driven nodeless oscillations (Bastrukov et all 2007b, 2008) whose general integral form is given by
 \begin{eqnarray}
  \label{e3.13}
  && {\cal K}_e= \frac{1}{2}\int \mu(r) [\nabla_i a_k({\bf r})+\nabla_k a_i({\bf r})]\,[\nabla_i a_k({\bf r})+\nabla_k a_i({\bf
 r})] \,d{\cal V}
 \end{eqnarray}
 where $\mu(r)$ stands for the shear modulus. Thus, in the
 standard two-component, core-crust, model of quaking neutron star (Franco et al 2000) presuming homogeneous density profile for
 crustal matter $\rho$ and constant values of
 transport coefficients of shear viscosity $\eta$ and shear elasticity $\mu$ one has
 $K_e(_0t_\ell)=(\mu/\eta){\cal D}(_0t_\ell)$;  the frequency of elastic nodeless vibrations
 is estimated by ratio $\omega^2={\cal K}_e/{\cal M}$ (Bastrukov et all 2007b, 2008).
 In the limit, $\lambda=(R_c/R)\to 0$, the analytic expression for the damping time, $\tau=2{\cal M}/{\cal D}$, of torsional nodeless oscillations in the crust is reduced to that for lifetime of global torsional nodeless oscillations in the entire volume of homogeneous neutron star model (Bastrukov et al 2003, 2007a)
\begin{eqnarray}
  \label{e3.14}
 && \tau(_0t_\ell)=\frac{2\tau_\nu}{(2\ell+3)(\ell-1)}.
 \end{eqnarray}
 The practical significance of outlined equations is that they can be used for the study of nodeless torsional vibrations of more wide class of celestial objects having solid seismogenic mantle of finite
 depth, such as Earth-like planets (Lay \& Wallace 1995, Aki \& Richards 2003) and white dwarf stars
 whose asteroseismology is the subject of intense current research (e.g. Winget \& Kepler 2008).

 Computation of integral for the spring constant ${\cal K}_m$ of the magnetic Lorentz restoring force of non-radial
 Alfv\'en nodeless differentially rotational vibrations in the crust yields
\begin{eqnarray}
 \label{e3.15}
 K_m&=&B^2R^3\frac{\ell(\ell^2-1)(\ell+1)}{(2\ell+1)(2\ell-1)}\,k_A^t(\ell,\lambda)\\
 \nonumber
 k_A^t(\ell,\lambda)&=&
 (1-\lambda^{2\ell+1})^{-2}\left\{1+\frac{3\lambda^{2\ell+1}}{(\ell^2-1)(2\ell+3)}\,
 \left[1-\frac{1}{3}\ell(\ell+2)(2\ell-1)
 \lambda^{2\ell+1}\right]\right\}.
 \end{eqnarray}
 As a result we arrive at the spectral equation for the frequency as a function of multipole degree of nodeless Alfv\'en oscillations
 \begin{eqnarray}
 \label{e3.16}
 && \nu^2(_0a^t_\ell[\nu_a,h])=\nu^2_A\left[(\ell^2-1)\frac{2\ell+3}{2\ell-1}\right]\,
 \frac{k_A^t(\ell,\lambda)}{m_t(\ell,\lambda)},\\
 \label{e3.15}
 && \omega_A=2\pi\nu_A=\frac{v_A}{R},\quad v_A=\frac{B}{\sqrt{4\pi\rho}},\quad
 \lambda=1-h,\,\, h=\frac{\Delta R}{R}
 \end{eqnarray}
 which depends, as one sees, upon two parameters -- Alfv\'en frequency, $\nu_A=(1/2\pi)v_A/R$ and the fractional
 depth, $h=\Delta R/R=1-\lambda$, of seismogenic layer.
 In Fig.4 and Fig.5, the fractional frequencies and periods of toroidal Alfv\'en vibration mode
 as a functions of multipole degree $\ell$ with indicated values of fractional depth of the seismogenic layer $h$
 are plotted, showing that the higher $\ell$ the larger is the frequency and the shorter is the period of Alfv\'en nodeless vibrations.

 \begin{figure}[h]
\centering{\includegraphics[width=9.cm]{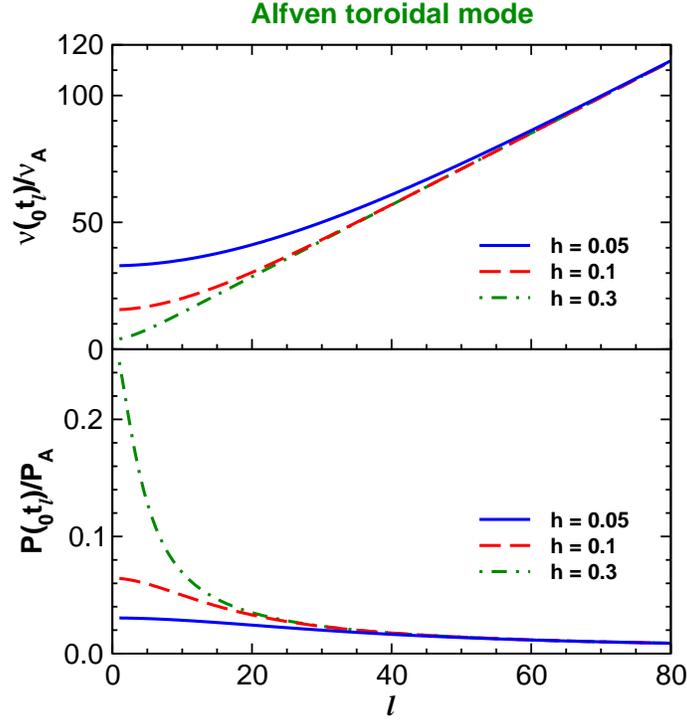}}
 \caption{Fractional frequency and period of nodeless torsional magneto-solid-mechanical oscillations,
 toroidal Alfv\'en mode -- $_0a^t_\ell$, entrapped in the neutron star crust as functions of multipole degree $\ell$
 at different values of the fractional depth $h$ of peripheral seismogenic crust. The value
 $h=1$ corresponds to global torsional oscillations excited in the entire volume of the star. Here $\nu_A=\omega_A/2\pi$, where $\omega_A=v_A/R$ with $v_A=B/\sqrt{4\pi\rho}$ being the velocity
 of Alfv\'en wave in crustal matter of density $\rho$ and $P_A=2\pi/\omega_A$.}
\end{figure}

\begin{figure}[h]
\centering{\includegraphics[width=9.cm]{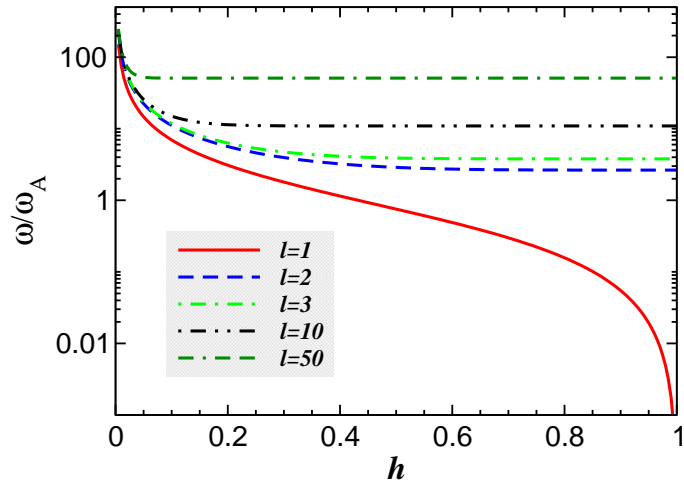}}
 \caption{Fractional frequency of nodeless torsional Alfv\'en oscillations
 of indicated overtone $\ell$ as a function of the fractional depth $h$ of peripheral seismogenic layer.
 The vanishing of dipole overtone in the limit of $h\to 1$, the case when entire mass of neutron star sets in torsional oscillations about dipole
 magnetic moment of the star value exhibits behavior which is typical for Goldstone's soft modes.}
\end{figure}
 One sees that the lowest overtone of torsion Alfv\'en oscillations entrapped in the crust is of dipole degree, $\ell=1$. However, in the limit $\lambda=(R_c/R)\to
 0$ [or, in other terms, when $h\to 1$ and $k_A^t(\ell,\lambda)/m_t(\ell,\lambda)\to 1$], the case when entire volume of the star sets in the oscillations, the above two-parametric spectral formula is reduced to the one-parametric frequency spectrum of global Alfv\'en torsional vibrations (Bastrukov et al 2009)
 \begin{eqnarray}
 \label{e3.17}
 && \nu(_0a^t_\ell)=\nu_A\left[(\ell^2-1)\frac{2\ell+3}{2\ell-1}\right]^{1/2},\,\,
 \nu_A=\frac{\omega_A}{2\pi},\quad\omega_A=\sqrt{\frac{RB^2}{3M}},\quad M=\frac{4\pi}{3}\rho\,R^3
 \end{eqnarray}
 with the lowest overtone of quadrupole degree, $\ell=2$. This suggests that dipole vibration is
 Goldstone's soft vibrational mode whose most conspicuous property is that the mode
 disappears (the frequency tends to zero) when one of parameters, in our case $\lambda$,
 of vibrating system turns into zero (Bastrukov et al 2008).

 Physically, the considered solid-mechanical model
 can be invoked to analysis of quasi-periodic oscillations of the
 electromagnetic emission from neutron star with approximately homogenous internal and dipolar external magnetic field
 when quake-induced perturbation is fairly week, that is,
 does not penetrate into the deep interior of the star setting in oscillations only its peripheral solid mantle of finite depth. Also, the model is appropriate for a case when the electrodynamic properties of the core material are incompatible with conditions of existence of Alfv\'en oscillations, one of the major of which is
 the extremely large (effectively infinite) electrical conductivity of stellar material (e.g., Chandrasekhar 1961, Mestel 1999). For the solid crustal matter, which is normally considered as a metal-like solid-state plasma
 composed of nuclei embedded in the degenerate Fermi-gas of relativistic electrons (whose pressure prevents the crust form gravitational contraction) this condition is certainly fulfilled (e.g. Chanmugham 1992).
 However, this is may not be the case for much denser matter of the neutron star cores, because its micro-composition
 is dominated, according to canonical notion of the neutron star, by degenerate Fermi-matter of non-relativistic neutrons (whose pressure is responsible
 for the neutron star gravitational equilibrium) -- the electrically neutral Fermi-particles
 endowed with intrinsic spin magnetic moments. This latter property of neutrons suggests that the
 degenerate Fermi-material of the neutron star cores can be brought to gravitational
 equilibrium in the magnetically ordered state of paramagnetic permanent magnetization caused by Pauli's mechanism of field-induced polarization of spin magnetic moment of neutrons along the axis of frozen in the core
 fossil magnetic field. It is just the case when the non-conducting core matter is capable of sustaining homogeneous
 magnetic field, but is unable to support Alfv\'en oscillations. The structure of such paramagnetic neutron star model, which is extensively discussed in (Bastrukov et al 2002a, 2002b, 2003) is schematically pictured in
 Fig. 6.

\begin{figure}[h]
\centering{\includegraphics[width=7.0cm]{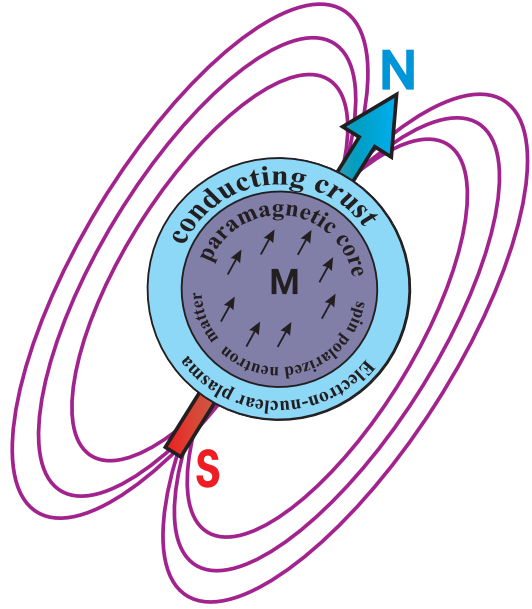}}
\caption{
 The internal constitution of two-component, core-crust, model of paramagnetic neutron star model.
 The core is considered as a poorly conducting permanent magnet composed of degenerate Fermi-gas of non-relativistic neutrons in the state of paramagnetic saturation caused by Pauli's paramagnetism -- field-induced alignment of spin magnetic moment of neutrons along the axis of uniform internal and dipolar external magnetic field frozen in the star on the stage of gravitational collapse of its massive progenitor. A highly conducting metal-like material of crust, composed of nuclei suspended in the degenerate Fermi-gas  of relativistic electrons, is regarded as magneto-active electron-nuclear solid-state plasma capable of sustaining Alfv\'en oscillations.}
\end{figure}

 It worth noting that internal constitution of such a star does not exclude the existence of a fairly thin inner
 crust which is predominately composed of boson matter whose micro-constituents are paired neutrons with average
 density several times less than the normal nuclear density, $0.1-0.5 \rho_N$.  Such a matter
 permits Bose-condensate state having some features in common with that generic to weakly compressed superfluid liquids. Understandably that such boson fraction of neutron matter, however, can only insufficiently contribute
 to the total mass budget of neutron star, because Bose-matter is unable to withstand the pressure of self-gravity, contrary to the degenerate neutron Fermi-matter whose pressure of degeneracy counterbalancing the self-gravity pressure is central to the very notion of neutron star. The physical importance of thin inner crust
 is, most likely, in that it operates like a lubricant facilitating
 differentially rotational shear of conducting crust relative to much denser paramagnetic core. From the view point of this model, the coupling between core and crust is one and the same of physical nature of magnetic
 cohesion as that operating between samples of terrestrial metal and permanent magnet which is mediated by the
 lines of magnetic force. The idea of decisive role of magnetic core-crust cohesion seems to
 be in line with the starquake model of the magnetar burst (Woods and Thompson 2006) which is though of as
 disruption of magnetic field lines in the process of explosive release of magnetic field stresses fracturing the neutron star crust.
 The presented line of argument (while admittedly speculative, but no more than any other considerations touching upon the internal structure of neutron stars) shows what inference can be made regarding electromagnetic properties of super dense stellar matter from asteroseismic model of Lorentz-force-driven torsional Alfv\'en nodeless oscillations entrapped in the finite-depth crust, provided of course that the above derived two-parametric spectral formula
 properly describe observational data on oscillating electromagnetic emission from quaking neutron star.
 In the next section we assess this conjecture by applying the derived theoretical spectrum to modal $\ell$-pole classification of QPOs in the X-ray flux during the flare of SGR 1806-20 and SGR 1900+14.

\section{QPOs in the X-ray flux during the flare of SGR 1806-20 and SGR 1900+14 from the viewpoint of above asteroseismic model}

  The physics behind interconnection between oscillating lines of magnetic field frozen in the star
  and quasi-periodic oscillations of detected electromagnetic flux has been recognized long ago (e.g. van Horn 1980, Blaes et al 1988) -- the quake-induced perturbation excites coupled
  oscillations of perfectly conducting matter and the lines of frozen-in magnetic field which
  outside the star operates as transmitters of beam of charged particles producing coherent radiation from the
  neutron star magnetosphere, so that the frequency of Alfv\'en seismic oscillations is considered to be
  equal to the frequency of quasi-periodic variations of the beam direction; this asteriseismic argument
  regarding physical nature of quasi-periodic oscillations is applicable to both radio emission from glitching pulsars and high-energy emission from bursting magnetars.

 \begin{figure}[h]
 \centering{\includegraphics[width=10.0cm]{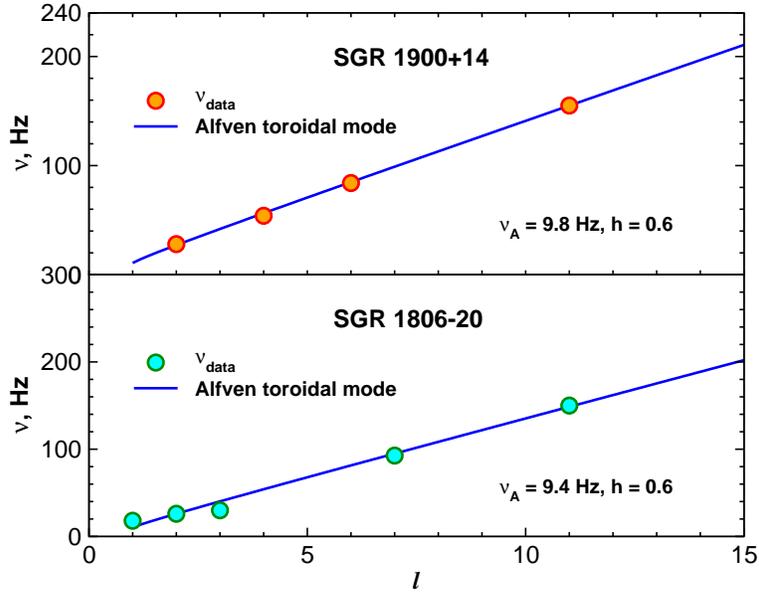}}
 \caption{Theoretical fit of the QPOs frequency in the X-ray flux during the flare of SGRs 1806-20 and 1900+14 by the obtained two-parametric spectral equation for the frequency of nodeless
 torsional magneto-solid-mechanical oscillations
 entrapped in the peripheral finite-depth crustal region of a neutron star with homogeneous internal and dipolar external magnetic field.}
\end{figure}

  As was noted in our previous study of global Alfv\'en oscillations in the entire volume of neutron star with homogeneous internal and dipolar external field (Bastrukov et al 2009), it is impossible reconcile
  one-parametric frequency spectrum $\nu(_0a^t_\ell)$ plotted as a function of multipole $\ell$, given by equation (\ref{e3.17}), with the observed QPOs frequencies (measured in Hz),
  which, according to compilation by Sotani et al (2008), for SGR 1806-20 are 18, 26, 29, 92, 150, 625, 1837, and for  SGR 1900+14 these are 28, 54, 84, 155. Bearing this in mind, one can attempt to
  fit the detected QPOs on the basis of derived here two-parametric spectral equation for torsional Alfv\'en nodeless oscillations entrapped in the finite-depth peripheral crust.
  The adjustable parameters of such a fit are the Alfv\'en frequency $\nu_A$
  and the fractional depth of seismogenic layer $h$. The suggested $\ell$-pole identification
  is summarized in Figs. 7 and 8.
  One sees that at indicated in this figures parameters,
  the model adequately regain the general tendency in the data on QPOs frequencies when
  $\ell$ are specified by integer numbers clearly recognizable in ticks of x-axis of Fig. 7.
  Deserved for special comment is that the low-frequency QPOs in data for SGR 1806-20, are fairly reasonably interpreted as dipole and quadrupole overtones of torsional magnetic Alfv\'en nodeless vibrations, namely, $\nu(_0a^t_1)=18$ and $\ell(_0a^t_2)=26$ Hz. The high-frequency kilohertz vibrations with $627$ Hz and $1870$ Hz in the QPOs spectrum
  of SGR 1806-20 highlighted in Fig.8
  are unambiguously specified as high-multipole overtones,
  namely, as $\nu(_0a^t_{\ell=42})=627$ Hz and $\nu(_0a^t_{\ell=122})=1870$ Hz.

\begin{figure}[h]
 \centering{\includegraphics[width=11.0cm]{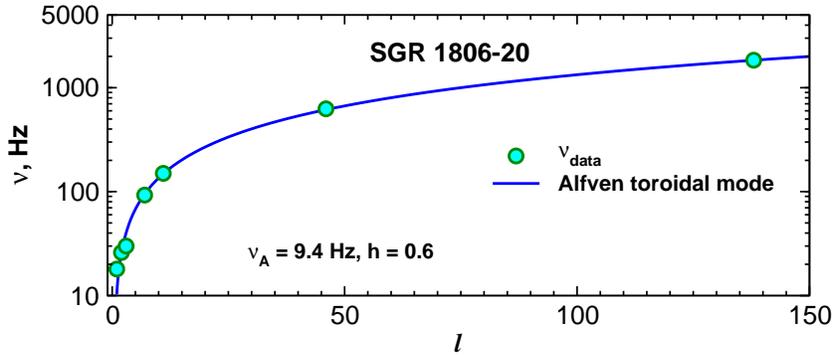}}
 \caption{The same as Fig.7, but for only SGRs 1806-20 with account for
 high-frequency QPOs.}
\end{figure}

  From the indicated in these figures values of $\nu_A=(1/2\pi)v_A/R$ it follows that
  the average speed of Alfv\'en wave $v_A$ in the crustal matter falls in the range $10^7 < v_A < 10^8$ cm sec$^{-1}$ and, thus, less than the average speed of elastic shear wave whose
  generally accepted value is $c_t\approx 10^8$  cm sec$^{-1}$. Also noteworthy is that the existence of free vibrations with frequency $\omega=2\pi\nu$ is determined by condition $\omega\tau >>1$, as follows from equation (\ref{e2.11}), which is too fulfilled; but this of course does not mean that the conversion of energy of Alfv\'en seismic crustal vibrations into heat (thermal emission) is the dominant mechanism of their exponential damping.
  All the above suggests that considered model of torsional nodeless Alfv\'en oscillations locked in the crust, provides somewhat better account of the data on the QPOs frequencies, as compared to the model
  of global torsional nodeless vibrations of magnetar under combined action of elastic Hooke's and magnetic Lorentz forces (Bastrukov et al 2009).

 \section{Summary}

 The neutron stars, both pulsars and magnetars, are seismically active compact objects of the end point of the evolutionary stellar track. They are, as is widely believed, to be formed in the magnetic-flux-conserving supernova collapse of the cores of massive stars coming into existence, as sources of pulsed electromagnetic emission,
 with a fairly complex internal constitution (e.g. Weber 1999, Bisnovatyi-Kogan 2002, Lattimer \& Prakash 2007).
 The past two decades have seen increasing recognition that valuable tool of looking into the neutron star interior
 provides asteroseimology. The explosive interest to current development of this domain of pulsar astrophysics has been stimulated by the above mentioned discovery of quasi-periodic oscillations during the giant X-ray flares of SGR 1806-20 and SGR 1900+14, which has also prompted extensive investigations of radiative processes caused by ultra-strong magnetic fields of magnetars
 (e.g. Heyl \& Hernquist 2005, Harding \& Lai 2006, Zhang et al 2007, Timokhin et al 2008, Bo Ma et al 2008, Rea et al 2008) and processes of their cooling (Yakovlev et al 2008).

 Many agree today that detected QPOs can be explained on the basis of asteroseismic models of a magnetar undergoing quake-induced torsional oscillations about its the dipole magnetic-moment axis. Adhering to this attitude and taking into account that the main energy source of bursting high-energy emission of magnetars is
 the immense store of the magnetic field energy, we have set up and investigated one of such models
 interpreting the above QPOs in terms of axisymmetric Alfv\'en nodeless oscillations of the crustal solid-state
 plasma about axis of homogeneous internal and dipolar external magnetic field frozen in the motionless
 core. As is demonstrated in Figs. 7 and 8, the obtained two-parametric spectral equation of this model
 provides proper account of general trends in data on QPOs frequencies
 as well as quite reasonable their $\ell$-pole identification. The main argument in favor of such interpretation
 is fairly accurate description of both the low-frequency (18 and 26 Hz) and high-frequency (627 and 1837 Hz)
  QPOs in X-ray flaring SGR 1806-20, which is presented in Fig.8, supporting the underlying this model conjecture about dominate role of Lorentz force of magnetic field stresses in maintaining the nodeless torsional oscillations of perfectly conducting solid-state plasma entrapped in the peripheral finite-depth seismogenic zone of magnetar.

  The physically meaningful framework for discussion of considered scenario of post-quake relaxation
  of magnetar provides the two-component, core-crust,
  model of quaking paramagnetic neutron star (Bastrukov et al 2002a, 2002b, 2003) which attributes the core matter incapability of supporting Alfv\'en differentially rotational shear oscillations to poor electrical
  conductivity of central region of the star, considering the magnetar core as a motionless spherical bar magnet
  composed of degenerate Fermi-matter of non-relativistic neutrons
  with spin magnetic moments polarized along the uniform fossil magnetic field frozen in the star
  on the stage of catastrophic contraction of its massive progenitor.
  Together with this, one cannot of course state that such an image of the quaking magnetar interior is unique, that is, that the above QPOs cannot be explained by alternative conjectures about electrodynamic properties of neutron star material, restoring forces and volumetric character of torsional oscillations (e.g. Bastrukov et al 2007b 2008, 2009). One more of such uninvestigated alternatives is the regime of nodeless
  global differentially rotational vibrations of neutron star with non-homogenous internal and dipolar external magnetic field presuming, thus, that stellar material in the total volume of magnetar possesses properties of perfect conductor.
  We shall return to such a model of quaking neutron star in a forthcoming paper.

\acknowledgments{This work is a part of projects on investigation of variability of high-energy emission from compact sources
 supported by NSC of Taiwan, grant numbers  NSC-96-2628-M-007-012-MY3 and NSC-97-2811-M-007-003.}

\end{document}